\documentclass{article}
\usepackage[pdftex]{graphicx}
\usepackage[cmex10]{amsmath}
\usepackage{amsfonts}
\usepackage{stfloats}
\usepackage{fullpage}
\usepackage{color}
\usepackage{amsmath,amsfonts,amssymb,hyperref}

\newcommand{\pd}[2]{\frac{\partial #1}{\partial #2}}
\newcommand{\pdd}[2]{\frac{\partial^2 #1}{\partial  #2^2}}

\newcommand{\vs}{v_{\rm s}}
\newcommand{\us}{u_{\rm s}}
\newcommand{\vsS}{v_{{\rm s}, noi}}
\newcommand{\usS}{u_{{\rm s}, noi}}

\newcommand{\uuf}{\mathbf{u}_{\rm f}}
\newcommand{\uue}{\mathbf{u}_{\rm e}}
\newcommand{\uus}{\mathbf{u}_{\rm s}}
\newcommand{\w}{\mathbf{w}}
\newcommand{\E}{\mathbf{E}}
\newcommand{\T}{\mathbf{T}}
\newcommand{\R}{\mathbb{R}}
\newcommand{\Tf}{\mathbf{T}_{\rm f}}
\newcommand{\boldS}{\mathbf{S}}
\newcommand{\tp}{^{\mathsf{T}}}
\newcommand{\mufr}{\mu_{\mathrm{fr}}}
\newcommand{\mue}{\mu_{\mathrm{e}}}
\newcommand{\lambdae}{\lambda_{\mathrm{e}}}
\newcommand{\rhoe}{\rho_{\mathrm{e}}}
\newcommand{\kappafr}{\kappa_{\mathrm{fr}}}
\newcommand{\kappaf}{\kappa_{\mathrm{f}}}
\newcommand{\kappas}{\kappa_{\mathrm{s}}}
\newcommand{\rhoa}{\rho_{\mathrm{a}}}
\newcommand{\rhos}{\rho_{\mathrm{s}}}
\newcommand{\rhof}{\rho_{\mathrm{f}}}
\newcommand{\cpI}{c_p^{\mathrm{I}}}
\newcommand{\cpII}{c_p^{\mathrm{II}}}
\DeclareMathOperator{\trace}{tr}

\begin{document}

\title{Deep convolutional neural networks for estimating porous
  material parameters with ultrasound tomography}

\author{Timo L\"ahivaara${}^a$, Leo K\"arkk\"ainen${}^b$, Janne M.J. Huttunen${}^b$, and Jan S. Hesthaven${}^c$} 
\date{${}^a$Department of Applied Physics,
  University of Eastern Finland, Kuopio, Finland\\\smallskip
${}^b$Nokia Technologies, Espoo,
 Finland\\Present address: Nokia Bell Labs, Espoo, Finland\\\smallskip
  ${}^c$Computational Mathematics and Simulation Science, Ecole
  Polytechnique F\'ed\'erale de Lausanne, Lausanne, Switzerland}

\maketitle

\subsection*{{\bf Abstract}}
  We study the feasibility of data based machine learning applied to
  ultrasound tomography to estimate water-saturated porous material
  parameters.  In this work, the data to train the neural networks is
  simulated by solving wave propagation in coupled
  poroviscoelastic-viscoelastic-acoustic media.  As the forward model,
  we consider a high-order discontinuous Galerkin method while deep
  convolutional neural networks are used to solve the parameter
  estimation problem.  In the numerical experiment, we estimate the
  material porosity and tortuosity while the remaining parameters
  which are of less interest are successfully marginalized in the
  neural networks-based inversion. Computational examples confirms the
  feasibility and accuracy of this approach.

\section{INTRODUCTION}

Measuring the porous properties of a medium is a demanding task.
Different parameters are often measured by different
application-specific methods, e.g., the porosity of rock is measured
by weighing water-saturated samples and comparing their weight with
dried samples. The flow resistivity of the damping materials can be
computed from the pressure drop caused to the gas flowing through the
material.  In medical ultrasound studies, the porosity of the bone is
determined indirectly by measuring the ultrasound attenuation as the
wave passes through the bone.

For the porous material characterization, information carried by waves
provides a potential way to estimate the corresponding material
parameters.  Ultrasound tomography (UST) is one technique that can be
used for material characterization purposes. In this technique, an
array of sensors is placed around the target. Typically, one of the
sensors is acting as a source while others are receiving the data.  By
changing the sensor that acts as a source, a comprehensive set of wave
data can be recorded which can be used to infer the material
properties.  For further details on the UST, we refer to
\cite{duric15} and references therein.

The theory of wave propagation in porous media was rigorously
formulated in 1950's and 1960's by Biot \cite{biot56a, biot56b,
  biot62a, biot62b}. The model was first used to study the porous
properties of bedrock in oil exploration. Since then, the model has
been applied and further developed in a number of different fields
\cite{sebaa06, yvonne14, lahivaara15, mart16}. The challenge of Biot's
model is its computational complexity. The model produces several
different types of waveforms, i.e., the fast and slow pressure waves
and the shear wave, the computational simulation of which is a
demanding task even for modern supercomputers. Computational
challenges further increase when attempting to solve inverse problems,
as the forward model has to be evaluated several times.

In this work, we consider a process for the parameter estimation,
comprising two sub-tasks: 1) Forward model: the simulation of the wave
fields for given parameter values and 2) Inverse problem: estimation
of the parameters from the measurements of wave fields.  The inverse
problem is solved using deep convolutional neural networks that
provide a framework to solve the inverse problems: we can train a
neural network as a model from wave fields to the parameters.  During
last decades, neural networks have been applied in various research
areas such as image recognition \cite{krichevsky,LeCun15}, cancer
diagnosis \cite{maclin, alvarez}, forest inventory
\cite{Muukkonen,niska}, and groundwater resources
\cite{Daliakopoulos2005229, jha10}.  Deep convolutional neural
networks (CNN) are a special type of deep neural networks
\cite{lecun98, Bengio09, LeCun15} that employ convolutions instead of
matrix multiplications (e.g. to reduce the number of unknowns).  In
this study, we employ deep convolutional neural networks to
two-dimensional ultrasound data.

The structure of the rest of the paper is as follows. First, in
Section \ref{sec:wavemod}, we formulate the poroviscoelastic and
viscoelastic models and describe the discontinuous Galerkin method.
Then, in Section \ref{sec:estim-mater-param}, we describe the neural
networks technique for the prediction of material parameters.
Numerical experiments are presented Section \ref{sec:numer-exper}.
Finally, conclusions are given in Section \ref{sec:conclusions}.

\subsection{Model justification}\label{sec:mmm}

\begin{figure}[!h]
\centering
\includegraphics[width=0.45\textwidth]{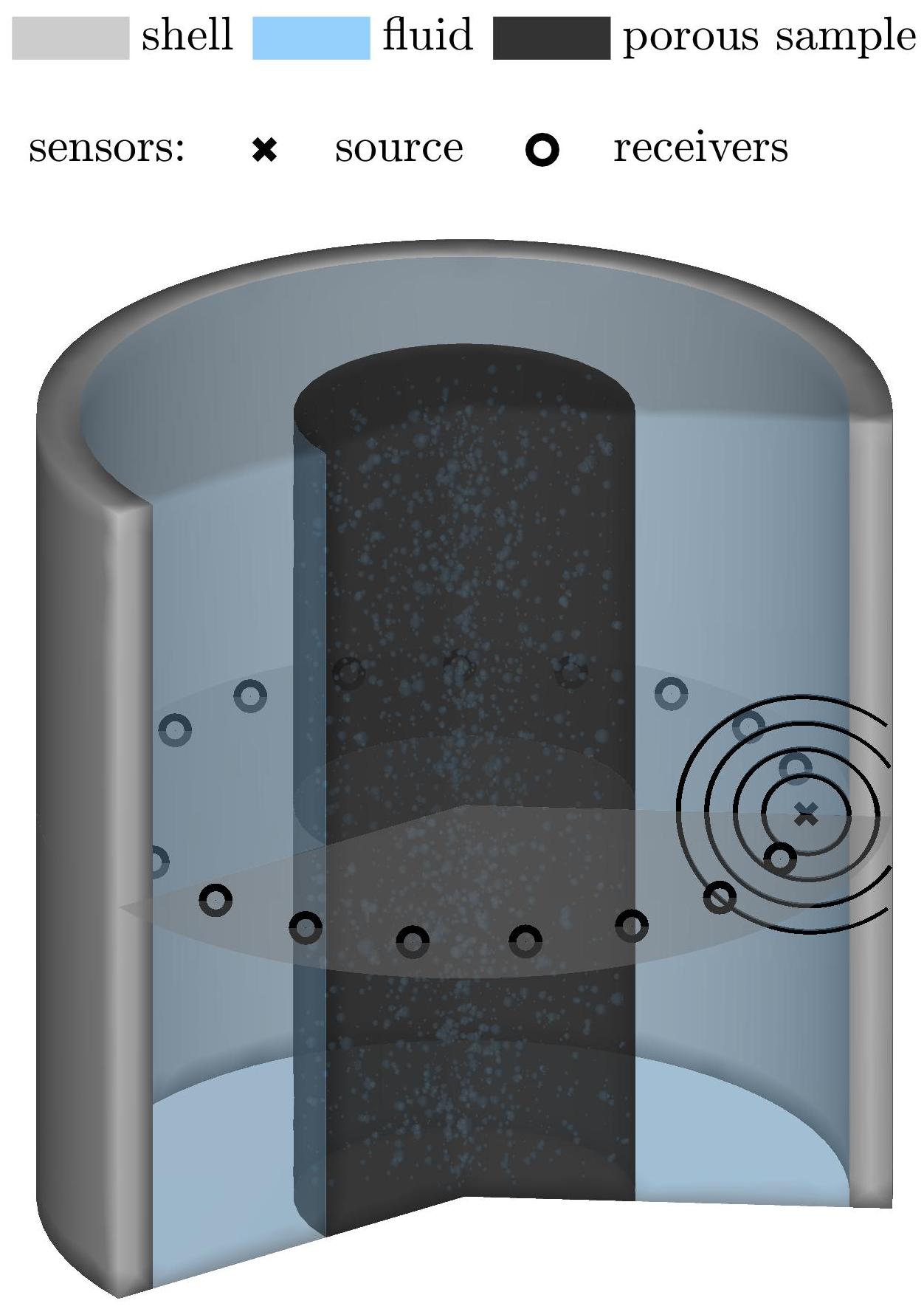}
\caption{\label{fig:geometry_3d} Schematic idea of the
  ultrasound tomography used to estimate porous material parameters.
  Acoustic wave generated by the source interact with the porous
  material which can be further seen on the data recovered by the
  receivers.}
\end{figure}

The purpose of this paper is to study the feasibility of using a data
based machine learning approach in UST to characterize porous material
parameters. In the synthetic model setup, we have a cylindrical water
tank including an elastic shell layer.  Ultrasound sensors are placed
inside the water tank.  In the model, we place a cylindrically shaped
porous material sample in water and estimate its porosity and
tortuosity from the corresponding ultrasound measurements by a
convolutional neural network. Figure \ref{fig:geometry_3d} shows a
schematic drawing of a UST setup studied in this work.

The proposed model has a wide range of potential applications.
Cylindrically shaped core samples can be taken of bedrock or manmade
materials, such as ceramics or concrete to investigate their
properties. In addition, from the medical point of view, core samples
can be taken, for example, of cartilage or bones for diagnosing
purposes. Depending on the application, the model geometry and sample
size together with sensor setup needs to be scaled.

\section{WAVE PROPAGATION MODEL}\label{sec:wavemod}

In this paper, we consider wave propagation in isotropic coupled
poroviscoelastic-viscoelastic-acoustic media.  In the following
Sections \ref{sec:poro} and \ref{sec:visc-atten}, we follow
\cite{carcione01,morency08,wilcox10,ward15} and formulate the
poroviscoelastic and viscoelastic wave models. Discontinuous Galerkin
method is discussed in Section \ref{sec:dg}.

\subsection{Biot's poroelastic wave equation}\label{sec:poro}

We use the theory of wave propagation in poroelastic media developed
by Biot in \cite{biot56a,biot56b,biot62a,biot62b}.  We express Biot
equations in terms of solid displacement $\uus$ and the relative
displacement of fluid $\w=\phi(\uuf -\uus )$, where $\phi$ is the
porosity and $\uuf$ is the fluid displacement. In the following,
$\rhos$ and $\rhof$ denote the solid and fluid densities,
respectively. We have
\begin{align}
  \label{eq:biot2a}
  \rhoa \pdd{\uus}{t} +\rhof \pdd{\w}{t} &=\nabla\cdot {\T},\\
  \rhof \pdd{\uus}{t} + \frac{\rhof \tau}{\phi} \pdd{\w}{t} + \frac{\eta}{k} \pd{\w}{t}\label{eq:biot2b}
                                         &=\nabla\cdot \Tf,
\end{align}
where $\rhoa= (1-\phi)\rhos + \phi \rhof$ is the average density, $\T$
is the total stress tensor, and $\Tf$ is the fluid stress tensor. In
Eq. (\ref{eq:biot2b}) $\tau$ is the tortuosity, $\eta$ is the fluid
viscosity, and $k$ is the permeability.

The third term in (\ref{eq:biot2b}) is a valid model at low
frequencies, when the flow regime is laminar (Poiseuille flow).  At
high frequencies, inertial forces may dominate the flow regime. In
this case, the attenuation model may be described in terms of viscous
relaxation mechanics as discussed, for example, in references
\cite{morency08,ward15}. In this paper, we use the model derived in
the reference \cite{ward15}. The level of attenuation is controlled by
quality factor $Q_0$.

One can express the stress tensors as
\begin{eqnarray}
  \T &=& 2\mufr {\bf E} + \left(\kappafr + \alpha^2 M -\frac{2}{3}\mufr\right)\trace({\bf E}){\bf I} 
  - \alpha M\zeta {\bf I}\label{eq:bfT},\\
  \Tf &=& M(\alpha\trace({\bf E}) - \zeta) {\bf I},\label{eq:bfTf}
\end{eqnarray}
where $\mufr$ is the frame shear modulus, $\E = \frac{1}{2}(\nabla
\uus +(\nabla\uus)\tp)$ denote the solid strain tensor, $\kappafr$ is
the frame bulk modulus, $\trace(\cdot)$ is the trace, {\bf I} is the
identity matrix, and $\zeta = -\nabla\cdot\w$ is the variation of the
fluid content.  The effective stress constant $\alpha$ and modulus
$M$, given in Eqs. (\ref{eq:bfT}) and (\ref{eq:bfTf}), can be written
as $\alpha = 1 - \kappafr/\kappas$, where $\kappas$ is the solid
bulk modulus, and $M =\kappas/\left(\alpha -
  \phi(1-\kappas/\kappaf)\right)$, where $\kappaf$ is the fluid bulk modulus.

\subsection{Viscoelastic wave equation}\label{sec:visc-atten}

The following discussion on the elastic wave equation with viscoelastic
effects follows Carcione's book \cite{carcione01}, in which a detailed
discussion can be found.  Expressed as a second order system, the elastic
wave equation can be written in the following form
\begin{equation}
\label{eq:viscoela}
  \rhoe\pdd{\uue}{t} = \nabla\cdot \boldS + {\bf s},
\end{equation}
where $\rhoe$ is the density, $\uue$ the elastic displacement,
$\boldS$ is a stress tensor, and ${\bf s}$ is a volume source.  In the
two-dimensional viscoelastic (isotropic) case considered here,
components of the solid stress tensor $\boldS$ may be written as
\cite{carcione01}
\begin{eqnarray}
\label{eq:sigma11}
  \sigma_{11} &=&
  (\lambdae+2\mue)\epsilon_{11}+\lambdae
  \epsilon_{22}+(\lambdae+\mue)\sum_{\ell=1}^{L_e}
  \nu_1^{(\ell)}+2\mue\sum_{\ell=1}^{L_e} \nu_{11}^{(\ell)},\\
\label{eq:sigma22}
  \sigma_{22} &=&
  (\lambdae+2\mue)\epsilon_{22}+\lambdae
  \epsilon_{11}+(\lambdae+\mue)\sum_{\ell=1}^{L_e}
  \nu_1^{(\ell)}-2\mue\sum_{\ell=1}^{L_e} \nu_{11}^{(\ell)},\\
\label{eq:sigma12}
  \sigma_{12} &=&
  2\mue \epsilon_{12}+2\mue\sum_{\ell=1}^{L_e} \nu_{12}^{(\ell)},
\end{eqnarray}
where $\mue$ and $\lambdae$ are the unrelaxed Lam{\'e} coefficients,
$\epsilon_{11}, \epsilon_{22},$ and $\epsilon_{12}$ are the strain
components, and $L_e$ is the number of relaxation terms.  The memory
variables $\nu_1^{(\ell)}$, $\nu_{11}^{(\ell)}$, and
$\nu_{12}^{(\ell)}$ satisfy
\begin{eqnarray}
  \frac{\partial \nu_1^{(\ell)}}{\partial t} &=& -\frac{\nu_1^{(\ell)}}{\tau_{\sigma\ell}^{(1)}}+\phi_{1\ell}(0)(\epsilon_{11}+\epsilon_{22}),\\
 \frac{\partial \nu_{11}^{(\ell)}}{\partial t} &=&
 -\frac{\nu_{11}^{(\ell)}}{\tau_{\sigma\ell}^{(2)}}+\frac{\phi_{2\ell}(0)(\epsilon_{11}-\epsilon_{22})}{2},\quad \ell=1,\ldots,L_e,\\ 
\frac{\partial \nu_{12}^{(\ell)}}{\partial t} &=& -\frac{\nu_{12}^{(\ell)}}{\tau_{\sigma\ell}^{(2)}}+\phi_{2\ell}(0)\epsilon_{12},
\end{eqnarray}
where
\begin{equation}
\label{eq:phiela}
  \phi_{k \ell}(t) = \frac{1}{\tau_{\sigma
      \ell}^{(k)}}\left(1-\frac{\tau_{\epsilon
        \ell}^{(k)}}{\tau_{\sigma
        \ell}^{(k)}}\right)\left(\sum_{\ell=1}^{n}\frac{\tau_{\epsilon\ell}^{(k)}}{\tau_{\sigma\ell}^{(k)}}\right)^{-1}\exp\left(-t/\tau_{\sigma l}^{(k)}\right)
  ,\quad  k=1, 2 .
\end{equation}
In Eq. (\ref{eq:phiela}), $\tau_{\epsilon \ell}^{(k)}$ and
$\tau_{\sigma \ell}^{(k)}$ are relaxation times corresponding to
dilatational $(k=1)$ and shear $(k=2)$ attenuation mechanisms.

Acoustic wave equation can be obtained from the system by setting the
Lam\'e coefficient $\mue$ to zero.

\subsection{Discontinuous Galerkin method}\label{sec:dg}

Wave propagation in coupled poroviscoelastic-viscoelastic-acoustic
media can be solved using the discontinuous Galerkin (DG) method (see
e.g. \cite{kaser06, puente08, wilcox10, gabard15, ward15}), which is a
well-known numerical approach to numerically solve differential
equations. The DG method has properties that makes it well-suited for
wave simulations, e.g., the method can be effectively parallelized and
it can handle complex geometries and, due to its discontinuous nature,
large discontinuities in the material parameters.  These are all
properties that are essential features for the method to be used in
complex wave problems.  Our formulation follows \cite{hesthavenbook},
where a detailed account of the DG method can be found.

\section{ESTIMATING MATERIAL PARAMETERS BY NEURAL NETWORKS}\label{sec:estim-mater-param}

The aim of this paper is to estimate porous material parameters by
applying artificial neural networks trained to simulated data.
Compared to traditional inverse methods, neural networks has an
advantage that it allows computationally efficient inferences. In
other words, after the network has been learned, inferences can be
carried out using the network without evaluating the forward
model. Furthermore, the neural networks provide a straightforward
approach to marginalize uninteresting parameters in the inference.

First we will give a brief summary to deep neural networks.  For a
wider representation of the topic, the reader is pointed to review
article by LeCun, Bengio, and Hinton \cite{LeCun15},
Bengio \cite{Bengio09} or to book by Buduma \cite{buduma17}.

We consider the following supervised learning task. We have a set of
input data $\{X_\ell\}$ with a labels $\{Y_\ell\}$ ($\ell=1,\ldots, N$).  
In our study, the input data comprises of the measured ultrasound
fields considered as ``images'' such that rows correspond to temporal
data and columns to receiver and the outputs are corresponding
material parameters.  The ``image'' should be interpreted as two
dimensional data, not a traditional picture; there is no color mapping
used in our algorithm.  The aim is to find a function $\Theta$ between
the inputs and outputs:
\begin{displaymath}
  Y=\Theta(X) .
\end{displaymath}
The task is to find a suitable form for the function and learn it from
the given data. Contrary to traditional machine leaning in which the
features are pre-specified, deep learning has an advantage that
features are learned from the data.

\begin{figure}[!h]
\centering
\includegraphics[width=0.8\textwidth]{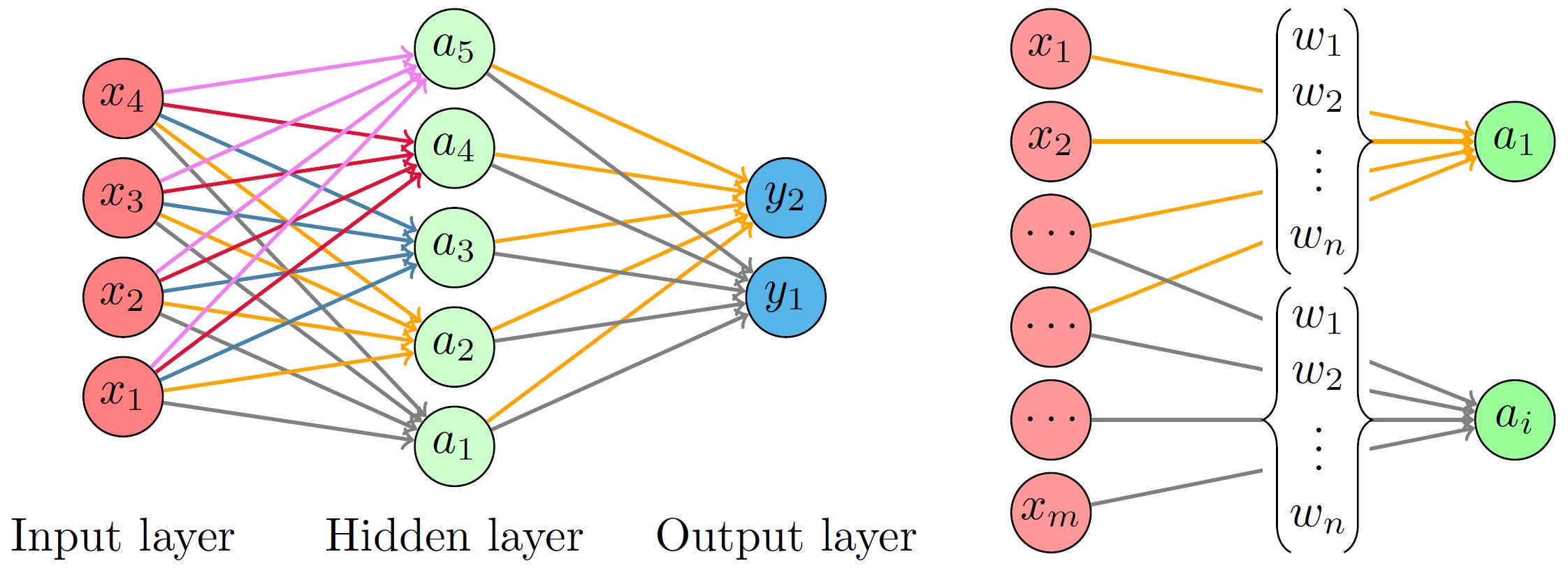}
\caption{\label{fig:networks} Left: Fully connected network with one
  hidden layer. Each arrow represents multiplication of the input or
  activation by a weight.  Right: A convolutional layer.
  $x_1,\ldots,x_m$ are the inputs to the layers and $a_1,\ldots,a_i$
  are the outputs of the layer. Note that the same set of the weights
  $w_1,\ldots,w_n$ are used for all outputs. }
\end{figure}

Neural networks is a widely used class of functions in machine
learning.  Figure \ref{fig:networks} (left) shows an example of a
neural network with one hidden layer.  Each circular node represents
an artificial neuron and an arrow represents a connection between the
output of one neuron to the input of the another.  This network has
four inputs $x_1, \ldots, x_4$ (input layer), five hidden units with
output activations $a_1,\ldots,a_5$ (hidden layer), and two outputs
$y_1$ and $y_2$ (output layer).  Each neuron in the hidden layer are
computational units that, for a given input $(x_1,\ldots,x_4)$,
calculates the activations as
\begin{displaymath}
  a_j=\sigma\left(\sum_{i=1}^4 w^{(1)}_{ji}x_i+b^{(1)}_j\right), \quad j=1,\ldots,5,
\end{displaymath}
where $w^{(1)}_{ji}$ is a weight assigned to the connection between
the $i$th input, $b^{(1)}$ is a bias terms and $j$th activation and
$\sigma$ is a non-linear function (nonlinearity).  Similarly, the
neurons at output layer are units for which the output is calculated
as
\begin{displaymath}
  y_j=\sigma\left(\sum_{i=1}^5 w^{(2)}_{ji}a_i+b^{(2)}_j\right) , \quad j=1,\ldots,2,
\end{displaymath}
where $w^{(2)}_{ji}$ and $b^{(2)}_j$ are again weights and biases.

At the present, the most typically used nonlinearity is the Rectified
Linear Unit (ReLU), $\sigma(x)=\max(0,x)$.\cite{hahnloser2000} During
past decades, smoother nonlinearities such as the sigmoid function
$\sigma(x)=(1+\exp(-x))^{-1}$ or tanh have been used as nonlinearity,
but the ReLU typically allows better performance in the training of
deep neural architectures on large and complex datasets.\cite{LeCun15}
Some layers can also have different linearities than other, but here
we use same $\sigma$ for notational convenience.

It is easy to see that the neural network can be presented in the
matrix form as
\begin{displaymath}
  A=\sigma(w^{(1)} \cdot X+b^{(1)}), \quad 
  Y=\sigma(w^{(2)} \cdot A+b^{(2)}), 
\end{displaymath}
where $X=(x_1,\ldots,x_4)^T$, $Y=(y_1,y_2)^T$, and
$A=(a_1,\ldots,a_5)^T$, $w^{(1)}$ and $w^{(2)}$ are
matrices/tensors comprising of the weights $w^{(1)}_{ij}$ and
$w^{(2)}_{ij}$, respectively, $b^{(1)}$ and $b^{(2)}$ are the vectors
comprising of the bias variables $b^{(1)}_{i}$ and $b^{(2)}_{i}$,
respectively, and the function $\sigma$ operates to the vectors
element-wise.  The network can be also written as a nested function
$\Theta_\theta (X)=\sigma\left(w^{(2)}\left(\sigma(w^{(1)} \cdot
    X+b^{(1)})\right)+b^{(2)}\right)$ where $\theta$ represents the
unknowns of the network ($\theta=(w^{(1)},b^{(1)},w^{(2)},b^{(2)})$).

Deeper networks can be constructed similarly.  For example, a network
with two hidden layers can be formed as
$A^{(1)}=\sigma(w^{(1)} \cdot X+b^{(1)})$,
$A^{(2)}=\sigma(w^{(2)} \cdot A^{(1)}+b^{(2)})$ and
$Y=\sigma(w^{(3)} \cdot A^{(2)}+b^{(3)})$, or
$\Theta_\theta(X)=\sigma\left(w^{(3)} \cdot\sigma\left(w^{(2)}
    \cdot\left(\sigma(w^{(1)} \cdot
      X+b^{(1)})\right)+b^{(2)}\right)+b^{(3)}\right)$ where
$\theta=(w^{(1)},b^{(1)},w^{(2)},b^{(2)},w^{(3)},b^{(3)})$.

The neural networks described above are called fully connected: there
are connections between all nodes of the network.  A disadvantage of
such fully connected networks is that total number of free parameters
can be enormous especially when the dimensions of input data is high
(e.g. high resolution pictures) and/or the network has several hidden
layers.  Convolutional neural networks (CNN) were
introduced \cite{fukushima80,lecun98} to overcome this problem.  Figure
\ref{fig:networks} (right) shows an example of a 1D-convolutional
layer.  In a convolutional layer, the matrix multiplication is
replaced with a convolution: for example, in 1D, the activations of
the layer are calculated as
\begin{displaymath}
  A=\sigma\left(w^{(1)} * X+B^{(1)}\right), 
\end{displaymath}
where $*$ denotes the discrete 1D-convolution and $w^{(1)}$ is a set
of filter weights (a feature to be learned).  In 2D, the input $X$ can
be considered as an image and the convolution $w^{(1)} * X$ is two
dimensional with a convolution matrix or mask $w^{(1)}$ (a filtering
kernel).

In practical solution, however, only one convolution mask (per layer)
is typically not enough for good performance.  Therefore, each layer
includes multiple masks that are applied simultaneously and the output
of the convolutional layer is a bank (channels) of images.  This three
dimensional object forms the input of the next layer such that each
input channel has own bank of filtering kernels and the convolution is
effectively three dimensional.

Usually convolutional neural networks employs also
pooling \cite{Scherer2010,krichevsky} for down-sampling (to reduce
dimension of activations).  See for example LeCun, Bengio, and
Hinton \cite{LeCun15} or Buduma \cite{buduma17} for details.

The training of the neural networks is based on a set of training data
$\{X_\ell,Y_\ell\}$. The purpose is to find a set if weights and
biases that minimize the discrepancy between the outputs $\{Y_\ell\}$
and the corresponding predicted values given by the neural networks
$\{\Theta_\theta(X_\ell) \}$.  Typically, in regression problems, this
is achieved by minimizing the quadratic loss function $f(\theta)$ over
the simulation dataset
\begin{equation}
  \label{eq:loss}
f(\theta)=f(\theta; \{X_\ell\}, \{Y_\ell\})=\frac{1}{N_{nn}}\sum_{\ell=1}^{N_{nn}} {
  \left(\Theta_\theta(X_\ell ) - Y_\ell \right)^2}
\end{equation}
to obtain the network parameters, weights, and biases, of the network.
The optimization problem could be solved using gradient descent
methods.  The gradient can be calculated using back propagation which
essentially is a procedure of applying the chain rule to the lost
function in an iterative backward fashion \cite{Linnainmaa70,
  buduma17,LeCun15}.  The computations of the predictions
$\Theta_\theta(X_\ell)$ and its gradients is computationally expensive
task in the case of large training dataset.  Therefore, during the
iteration, the cost and gradients are calculated for mini-batches that
are randomly chosen from the training set.  This procedure is called
as stochastic gradient descent (SGD) \cite{lecun98,buduma17}.

The validation of the network is commonly carried out using a test
set, which is either split from the original data or collected
separately.  This test set is used to carry final evolution of the
performance.  This is done due to the tendency of deep networks for
overfitting, which comes out as a good performance in the training set
but very poor performance in test set (i.e. the network ``remembers''
the training samples instead of learning to generalize).  Third set,
the validation data set, is also traditionally used to evaluate
performance during the training process.

\color{black}
\section{NUMERICAL EXPERIMENT}\label{sec:numer-exper}

In this section, we present the results obtained from testing the data
driven approach to estimate porous material parameters in ultrasound
tomography. In this paper, initial conditions are always assumed to be
zero and we apply the free condition as a boundary condition.

In the following results, time integration is carried out using an
explicit low-storage Runge-Kutta scheme \cite{carpenter94}. For each
simulation, the length of the time step $\Delta t$ is computed from
\begin{equation}
  \label{eq:CFL}
  \Delta t =
  \left(\frac{h^{\ell}_{\min}}{2c^{\ell}_{\max}(N^{\ell})^2}\right)_{\min}, \quad \ell=1,\ldots,K,
\end{equation}
where $c^{\ell}_{\max}$ is the maximum wave speed, $N^{\ell}$ is the
order of the polynomial basis, $h^{\ell}_{\min}$ is the smallest
distance between two vertices in the element $\ell$, and $K$ is the
number of elements. 

\subsection{Model setup}

\begin{figure}[!h]
\centering
\includegraphics[width=0.45\textwidth]{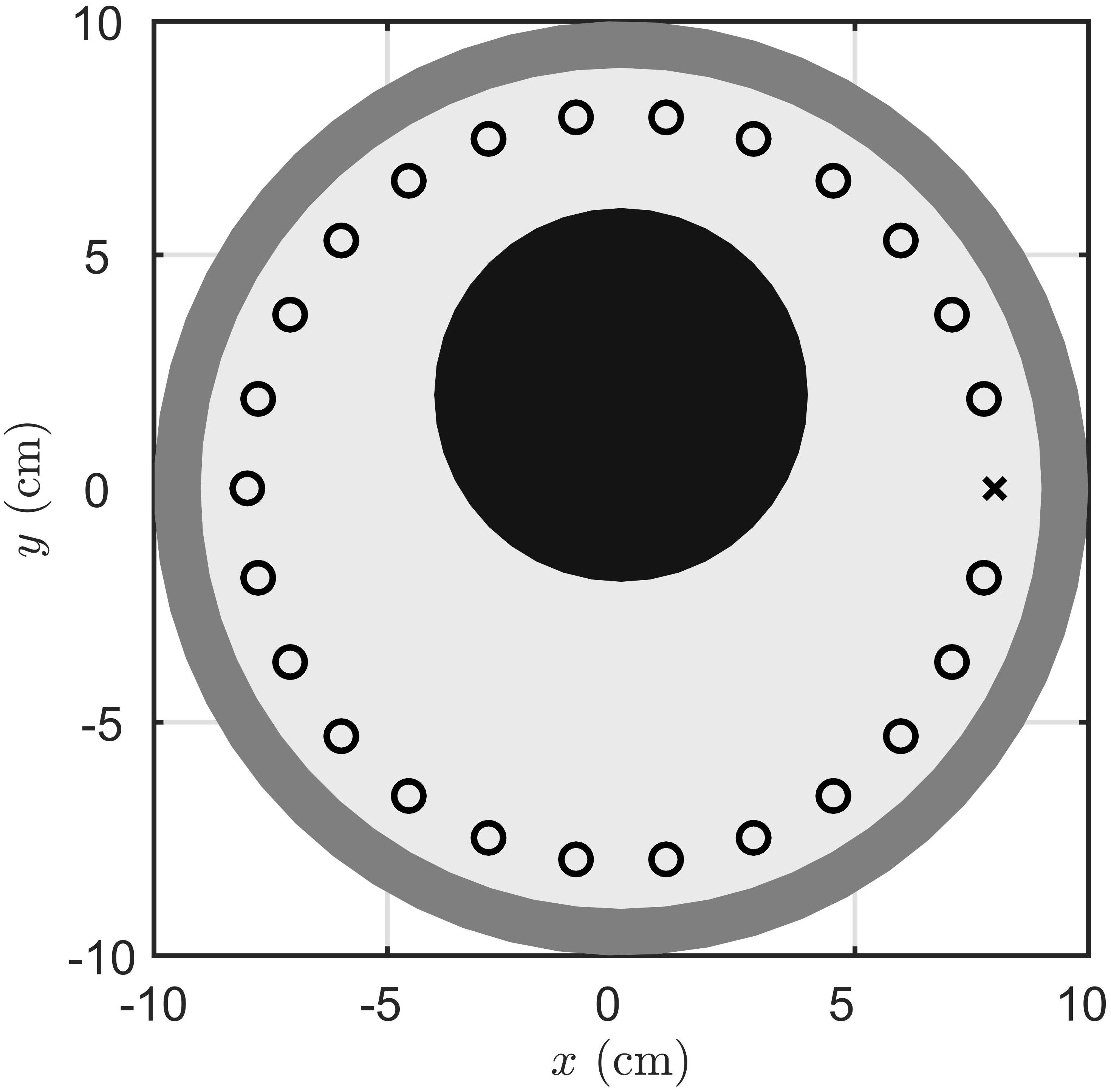}
\caption{\label{fig:geometry}{Figure shows the problem geometry. In
    the graph, the circles denote the receivers and the cross denotes
    the source.}}
\end{figure}

Let us first introduce the model problem. Figure \ref{fig:geometry}
shows the studied two dimensional problem geometry.  The propagation
medium contains three subdomains: a cylindrical shaped poroelastic
inclusion (black), a fluid (light gray), and a solid shell (dark
gray).  The computational domain is a circle with radius of 10 cm with
a 1 cm thick shell layer.  A circular shaped poroelastic inclusion
with a radius of 4 cm is vertically shifted to 2 cm from the center of
the circular domain.  Shifting the target from the center avoids
symmetrically positioned sensors (with respect to x-axis) to receive
equal signal and therefore have more information from the target.

A total of 26 uniformly distributed ultrasound sensors are located at
a distance of 8 cm from the center of the circle.  The source is
introduced on the strain components $\epsilon_{11}$ and
$\epsilon_{22}$ by the first derivative of a Gaussian function with
frequency $f_0=40$ kHz, a time delay $t_0=1.2/f_0$. The sensor that is
used as a source does not collect data in our simulations. Receivers
collect solid velocity components $\us$ and $\vs$. In the following,
the simulation time is 0.4 ms. Note that recorded data is downsampled
to a sampling frequency of 800 kHz on each receiver.

The inclusion is fully saturated with water. The fluid parameters are
given by: the density is $\rhof=1020$ kg/m$^{3}$, the fluid bulk
modulus is $\kappaf=2.295$ GPa, and the viscosity is $\eta=1.0$e-3
Pa$\cdot$s.  All other material parameters of the inclusion are
assumed to be unknown. In this paper, we assume a relatively wide
range of possible parameter combinations, see Table \ref{tab:bounds}.
Furthermore, the unknown parameters are assumed to be uncorrelated.
The physical parameter space gives $\sim$1.5 kHz as an upper bound for
Biot's characteristic frequency
\begin{equation}
  \label{eq:fc}
  f_c = \frac{\eta\phi}{2\pi \tau\rhof k}
\end{equation}
and hence we operate in Biot's high-frequency
regime $(f_c < f_0)$ in all possible parameter combinations.

\begin{table}[!h]
\centering
  \caption{Table lists the minimum and maximum values used for uniform
    distributions for each unknown physical
    parameter.}\label{tab:bounds}
  \begin{tabular}{cc|cc}
    \textbf{variable name} & \textbf{symbol}  & \textbf{minimum} & \textbf{maximum}\\
    \hline
solid density&$\rhos$ (kg/m$^{3}$) & 1000.0 & 5000.0\\
solid bulk modulus&$\kappas$ (GPa) & 15.0 & 70.0\\
frame bulk modulus&$\kappafr$ (GPa) & 5.0 & 20.0\\
frame shear modulus&$\mufr$ (GPa) & 3.0 & 14.0\\
tortuosity&$\tau$  & 1.0 & 4.0\\
porosity&$\phi$ & 0.01 & 0.99\\
permeability&$k$ (m$^2$)   & 1.0e-10 & 1.0e-7\\
quality factor&$Q_0$ & 20.0 & 150.0\\
  \end{tabular}
\end{table}

For the fluid subdomain (water), we set: the density $\rhoe=1020$
kg/m$^{3}$, the first Lam\'e parameter $\lambdae=2.295$ GPa, and the
second Lam\'e parameter $\mue = 0$. The elastic shell layer has the
following parameters: $\rho_e=2000$ kg/m$^{3}$, $\lambda_e=12.940$
GPa, and $\mu_e=5.78$ GPa.  The relaxation times for the viscoelastic
attenuation in the shell layer are given in Table \ref{tab:relax}.

\begin{table}[!h]
\centering
\caption{Table lists the relaxation times (two mechanisms) used with the shell layer. Relaxation times are computed using the nonlinear optimization method discussed in detail in \cite{Blanc01042016}. }\label{tab:relax}
    \begin{tabular}{c|cccc}
      & $\tau_{\epsilon \ell}^{(1)}$ & $\tau_{\sigma \ell}^{(1)}$ &  $\tau_{\epsilon \ell}^{(2)}$ & $\tau_{\sigma \ell}^{(2)}$\\
    \hline
    $\ell=1$ & 1.239e-4 & 1.176e-4 & 1.249e-4 & 1.165e-4\\
    $\ell=2$ & 5.319e-6 & 5.042e-6 & 5.370e-6 & 4.999e-6\\
  \end{tabular}
\end{table}

The derived wave speeds for each subdomain are given in Table
\ref{tab:derived_hete2}. For the poroelastic inclusion, both the
minimum and maximum wave speeds are reported. It should be noted that
the reported values for the wave speeds in the inclusion correspond to
the values generated by sampling the material parameters and hence
they may not correspond to the global maximum and minimum values. A
detailed approach for calculating wave speeds is given in
\cite{ward15}.

\begin{table}[!h]
\centering
  \caption{Derived wave speeds for each subdomain.  For the inclusion, both the
    minimum/maximum values are given. For the inclusion, values are
    based on sampling the material
    parameters.}\label{tab:derived_hete2}
      \begin{tabular}{c|ccc}
        {\bf subdomain}  &  $\cpI$ (m/s)& $\cpII$ (m/s)& $c_s$ (m/s)\\
        \hline
        {\bf inclusion} &  1900/20189 &196/1983 &833/11578\\
        {\bf fluid}      & 1500   & -  &  -\\
        {\bf shell}      & 3500   & -  &  1700\\
      \end{tabular}
\end{table}

It should be noted that, depending on the application, only some of
the material parameters are unknowns of interest. In this work, we
focus on estimating the porosity and tortuosity of the inclusion while
the solid density and bulk modulus, frame bulk and shear modulus,
permeability, and quality factor are marginalized in the neural
networks-based inversion algorithm, discussed in detail below.

\subsection{Training, validation, and test data}\label{sec:train-data-gener}

For the convolutional neural networks algorithm used in this work, the
recorded wave data is as expressed as images $X\in \R ^{d},\ d=N_t
\times N_r$, where $N_t$ denotes the number of time steps and $N_r$
the number of receivers. The input dimension is $d=320 \times 25$, as
the data $X$ can be seen as a 2D-image, comprising 25 pixels,
corresponding to the receiver positions, times 320 pixels
corresponding to the time evolution of the signal.

\begin{figure}[!h]
\centering
\includegraphics[width=0.45\textwidth]{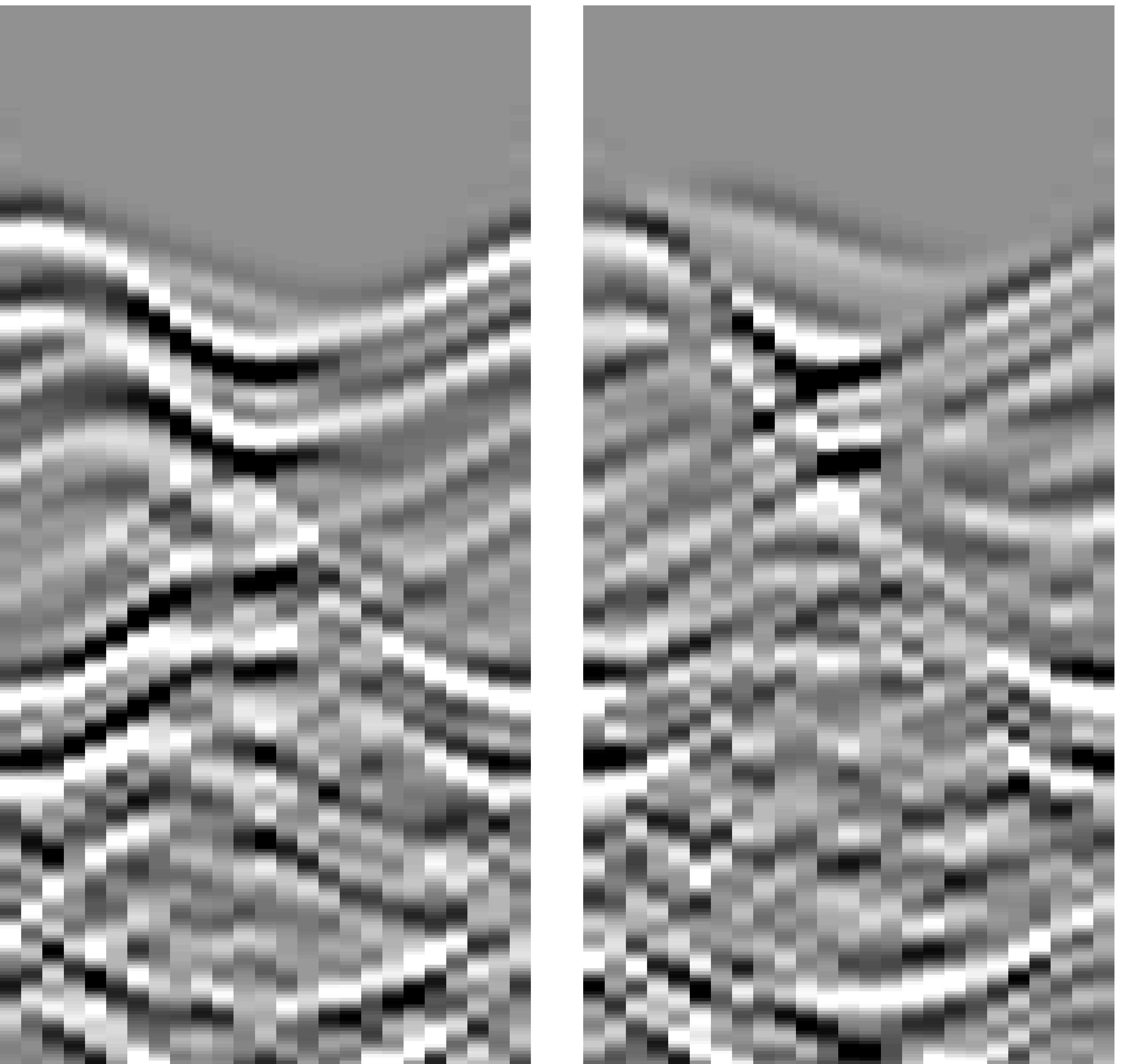}
\caption{Example images used for training the neural networks
  algorithm. Note that neural networks algorithm uses the original
  pixel values of the image $X$ (see Eq. (\ref{eq:normal})). Time is
  shown on the vertical-axis and receiver index on the
  horizontal-axis. Images correspond to samples where the shear wave
  speeds $c_s$ is minimum (left) or maximum (right) (see Table
  \ref{tab:derived_hete2}).}
  \label{fig:signals_teaching}
\end{figure}

As the data on each column of the image, we use the following
\begin{equation}
  \label{eq:normal}
 X(:, \ell) = \frac{x_r^{\ell}(\us^{\ell}-\usS^{\ell})+y_r^{\ell}(\vs^{\ell}-\vsS^{\ell})}{\sqrt{(x_r^{\ell})^2+(y_r^{\ell})^2}},\quad\ell=1,\ldots,N_r,
\end{equation}
where $(x_r^{\ell},\ y_r^{\ell})$ are the coordinates of the $\ell$th
receiver and $(\us^{\ell},\ \vs^{\ell})$ are the horizontal and
vertical velocity components on the $\ell$th receiver. In Eq.
(\ref{eq:normal}) subscript $noi$ denotes the data that is simulated
without the porous inclusion. Two example images are shown in Fig.
\ref{fig:signals_teaching}.  

We have generated a training data set comprising 15,000 samples using
computational grids that have $\sim$3 elements per wavelength. The
physical parameters for each sample are drawn from the uniform
distribution (bounds given in Table \ref{tab:bounds}).  The order of
the basis functions is selected separately for each element of the
grid. The order $N_{\ell}$ of the basis function in element $\ell$ is
defined by
\begin{equation}
\label{eq:porder}
  N_{\ell} = \left\lceil \frac{2\pi
  ah^{\ell}_{\max}}{\lambda^w_{\ell}}+b\right\rceil, 
\end{equation}
where $\lambda^w_{\ell}=c^{\ell}_{\min}/f_0$ is the wavelength,
$c^{\ell}_{\min}$ is the minimum wave speed, and $\lceil\cdot\rceil$
is the ceiling function.  The parameters $a$ and $b$ control the local
accuracy on each element.  Following \cite{lahivaara11, lahivaara10},
we set $(a,\ b)=(1.0294,\ 0.7857)$.

\begin{figure}[!h]
\centering
\includegraphics[width=0.47\textwidth]{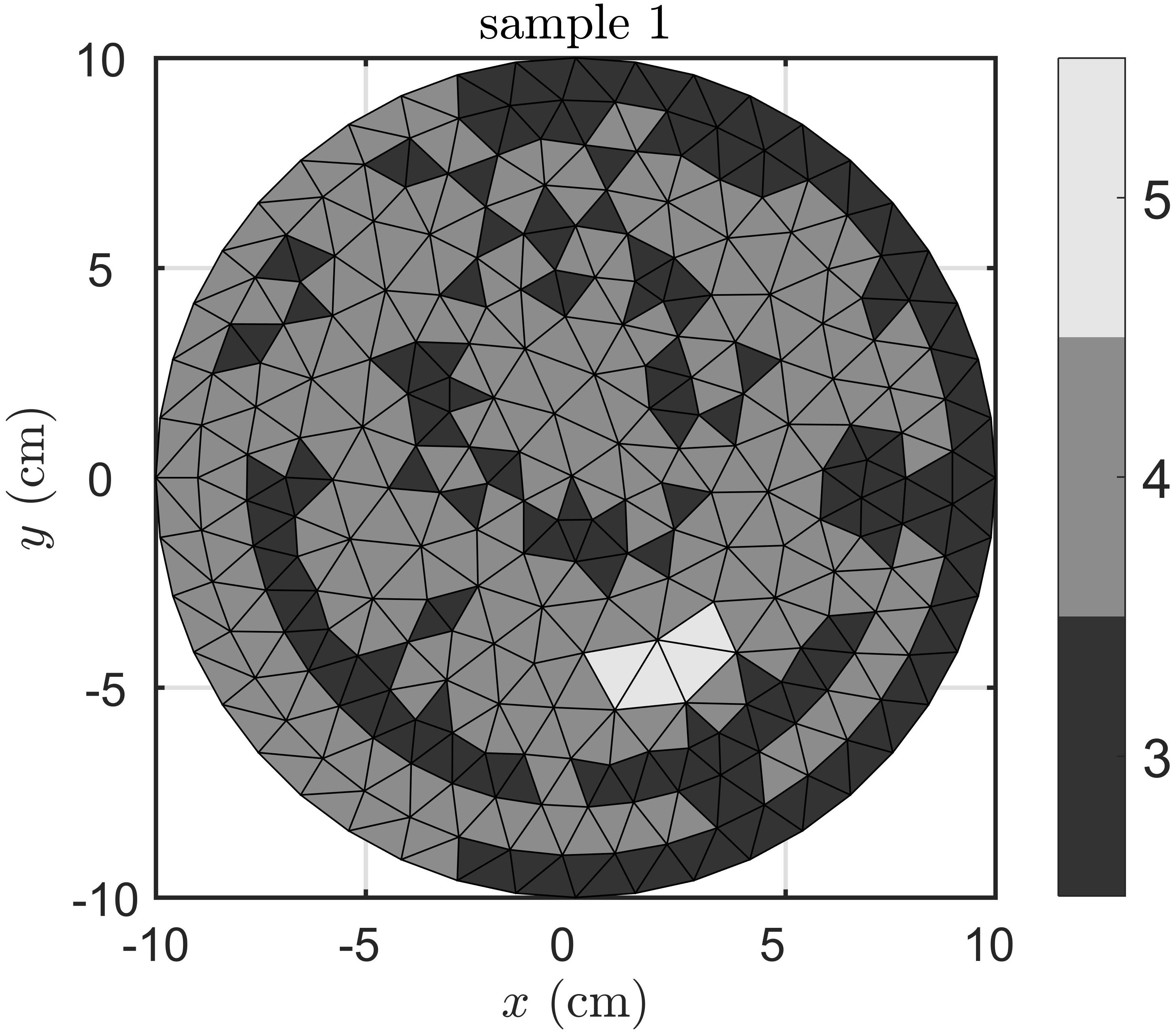}
\includegraphics[width=0.47\textwidth]{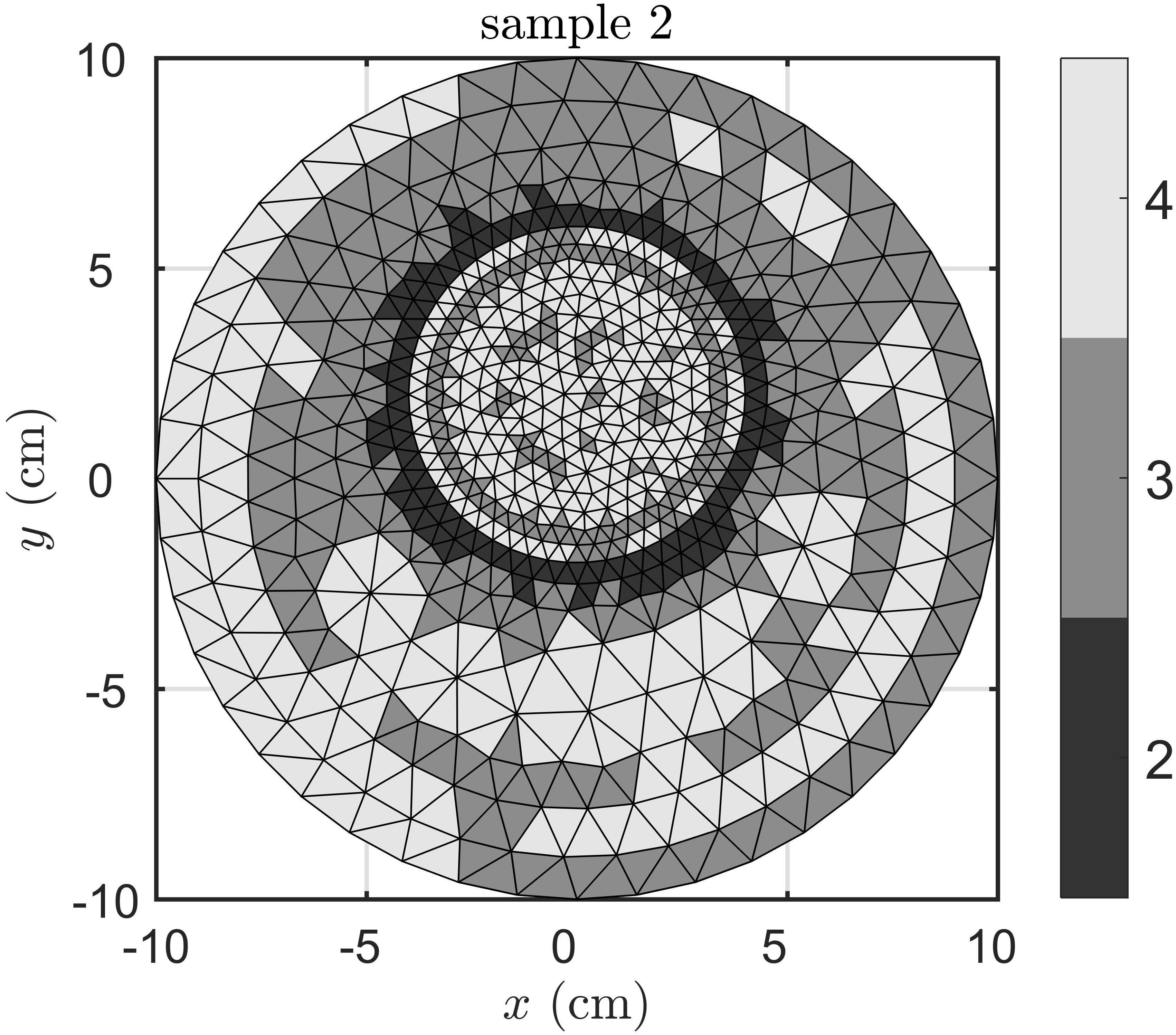}
\caption{Figure shows two example grids used in the computations. The
  colorbar shows the order of the basis functions. }
\label{fig:mesh_teaching}
\end{figure}

Figure \ref{fig:mesh_teaching} shows two examples of computational
grids and the corresponding basis order selection on each triangle.
Example grids consist of 466 elements and 256 vertices
($h_{\min}=0.81$ cm and $h_{\max}=1.95$ cm) (sample 1) and 1156
elements and 601 vertices ($h_{\min}=0.32$ cm and $h_{\max}=1.75$ cm)
(sample 2).

\begin{figure}[!h]
\centering
\includegraphics[width=0.47\textwidth]{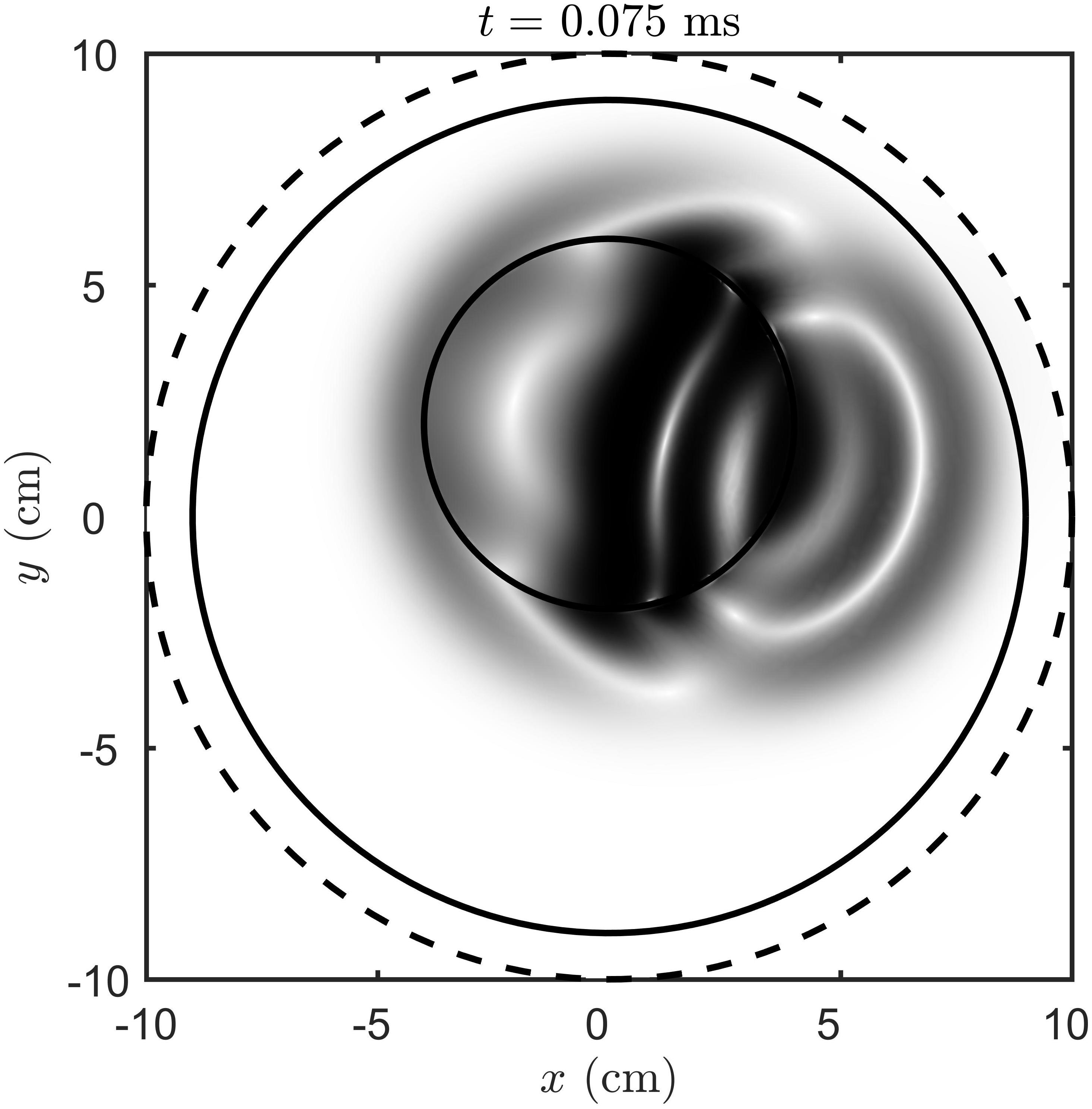}
\includegraphics[width=0.47\textwidth]{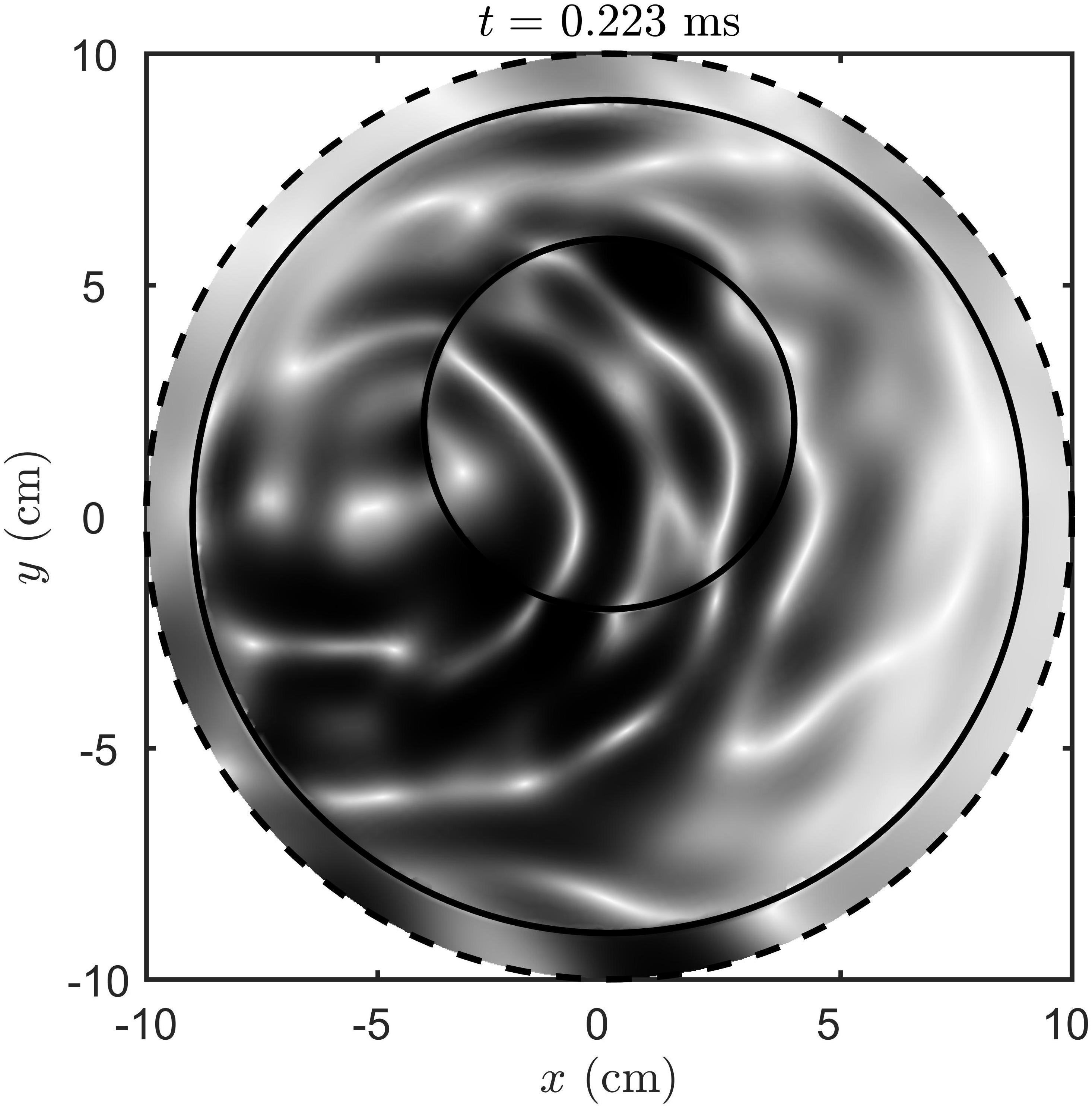}
\caption{Wave interaction with the poroelastic material at two times.
  The title shows the time. In the graphs, solid black lines show the
  inclusion/water and water/shell interfaces while the dotted black
  line shows the exterior boundary.}
  \label{fig:snaps}
\end{figure}

Figure \ref{fig:snaps} shows two snapshots of the scattered solid
velocity field ($\sqrt{(\us-\usS)^2+(\vs-\vsS)^2}$) for two time
instants. At the first time instant, the transmitted fast pressure
wave and also the first reflected wave front is clearly visible. At
the second time, all wave components have reflected back from the left
surface of the phantom. Furthermore, more complicated wave scattering
patterns can be seen inside the porous inclusion. These wave fields
demonstrate how the received signals are obtained as combinations of
multiple wave fronts.

To include observation noise in the training, each image in the
simulation set is copied 5 times and each of the copies is corrupted
with Gaussian noise of the form
\begin{equation}\label{eq:noisemodel}
  X_{\ell'}^\textrm{noised}=X_\ell+A\epsilon^A+B|X_\ell|\epsilon^B,
\end{equation}
where $\epsilon^A$ and $\epsilon^B$ are independent zero-mean
identically distributed Gaussian random variables.  The second term
represents additive white noise and the last term represents noise
relative the signal strength.  To represent a wide range of different
noise levels, for each sample image, the coefficients $A$ and $B$ are
randomly chosen such that the standard deviations of the white noise
component is between 0.03-5\% (varying logarithmically), and the
standard deviations of the relative component is between 0-5\%.  The
total number of samples in the training set is $N_{nn}=5\times
15000=75000$.

Furthermore, two additional data sets were generated: a validation
data set and a test set, that both comprise 3000 samples. In machine
learning, the validation data set is traditionally used to evaluate
performance during the training process and the test set is used for
the final evaluation of the network. These data sets are generated
similarly as the training set, except computational grids were
required to have $\sim$4 elements per wavelength to avoid inverse
crime \cite{kaipio07}. Furthermore, for the test set, the non-uniform
basis order parameters are $(a,\ b)=(1.2768,\ 1.4384)$
\cite{lahivaara11} (Eq. (\ref{eq:porder})) and the noise is added in a
more systematic manner (instead of choosing $A$ and $B$ are randomly)
to study the performance with different noise levels (see Results
section).

\subsection{Convolutional neural networks architecture}\label{sec:CNN}

\begin{figure}[!h]
\centering
\includegraphics[width=0.9\textwidth]{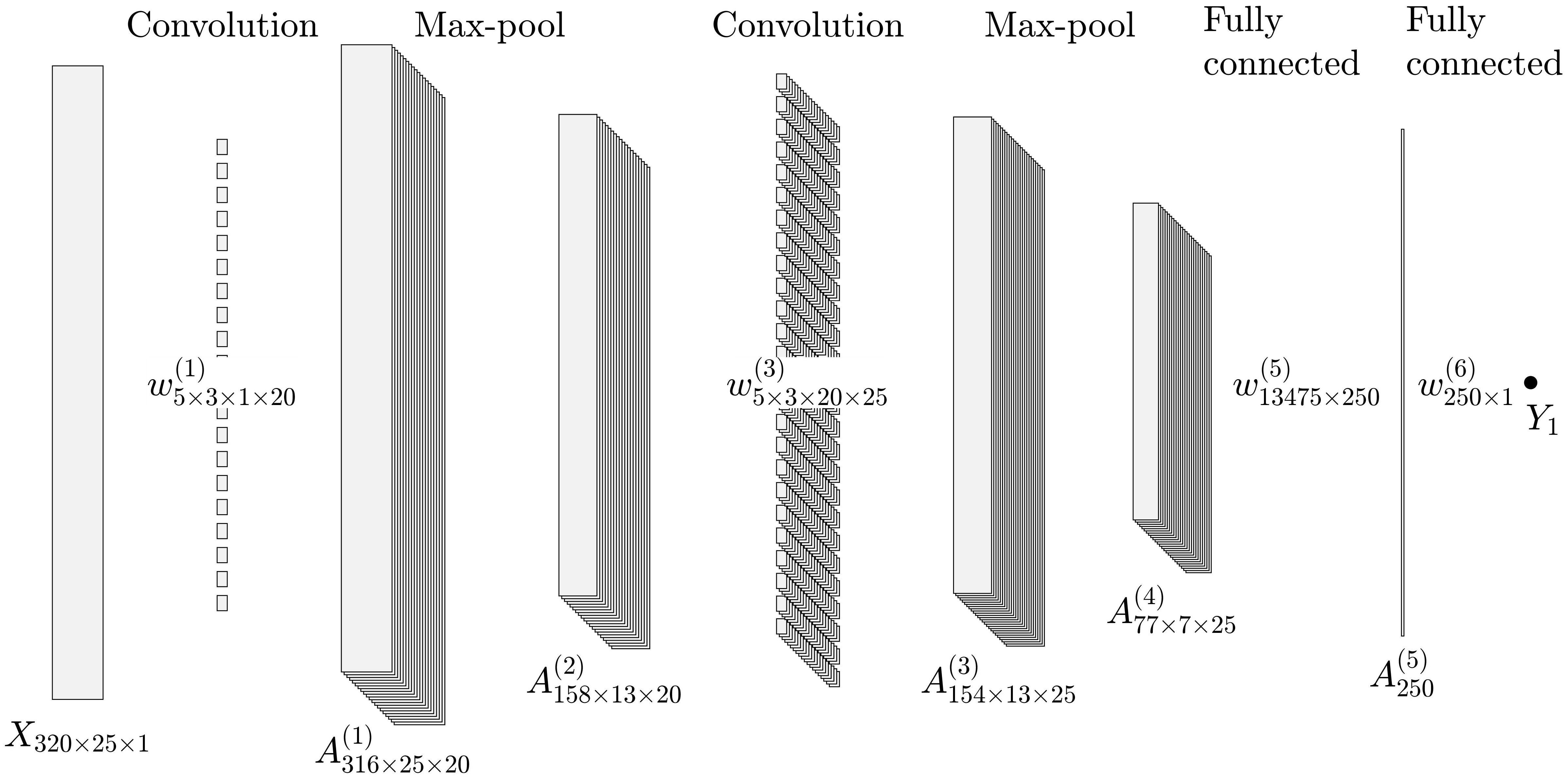}
\caption{A flow diagram of the network used in the study. Flow is from
  left to right (connecting arrows are not shown for graphical
  clarity). The sizes of the boxes scale with the memory size of the
  elements. The size/structure of the neurons and weights are also
  displayed in the subscript of $A^{(i)}$ and $w^{(i)}$ for each layer
  $i$. The matrices of the weights of the full connected layers
  $w^{(5)}$ and $w^{(6)}$ are not drawn for graphical clarity.}
  \label{fig:architecture}
\end{figure}

The neural network architecture is shown in Table
\ref{tab:network} and Fig. \ref{fig:architecture}.  The CNN
architecture is similar to ones used for image classification, for
example, Alexnet \cite{krichevsky}, with some exceptions.  In our
problem, the input data can be considered as one channel images
instead of color images with three channels (i.e. we do not apply any
color mapping).  Our network also lacks the softmax layer, which is
used as an outmost layer to provide the classification.  Instead, the
outmost two layers are simply fully connected layers. In addition, our
network also has smaller number of convolutional and fully connected
layers with and smaller dimensions in filter banks, which leads to
significantly smaller number of unknowns.  We, however, wish to note
that purpose aim was not to find the most optimal network
architecture.  For example, there can be other architectures that can
provide similar performance with even smaller number of the unknowns.
Furthermore, similar performance can also be achieved with fully
connected networks with $\sim$3 layers, but at the expense of
significantly larger number of unknowns.

The loss function is chosen to be quadratic (Eq.
(\ref{eq:loss})).  The implementation is carried out using Tensorflow,
which is a Python toolbox for machine learning.  The optimization was
carried out with the Adam optimizer \cite{Kingma2014}.  The batch size
for the stochastic optimization is chosen to be 50 samples.

\begin{table}[!h]
\centering
\caption{The convolutional neural network architecture used in this
  work.  The convolutional layers uses periodic convolution (padding)
  in the receiver position-direction and no padding in the temporal direction. 
  The total number of unknowns in the network is 3.38M. }\label{tab:network}
  \begin{tabular}{|c|c|c|c|}
     \hline
    \hline
    \textbf{Layer k} &  \textbf{Type and non-linearity} & \textbf{input size} & \textbf{output size}\\
    \hline
                     & Input &  $320 \times 25$ &  $320 \times 25
                                                  \times 1$\\   
    \hline
    1 & Convolution layer ($5\times 3\times 1$ filter, 20 filters)  &  $320 \times 25 \times 1$& $316 \times 25 \times 20$\\
     & + Layer normalization + ReLU & &\\
     & + Max-pooling ($2\times 2$) &  $316 \times 25 \times 20$& $158 \times 13 \times 20$\\
    2 & Convolution layer ($5\times 3\times 20$ filter, 25 filters) &  $158 \times 13 \times 20$& $154 \times 13 \times 25$\\
     & + Layer normalization + ReLU & &
\\
     & + Max-pooling ($2\times 2$) &  $154 \times 13 \times 25$& $77 \times 7 \times 25$\\
    3 & Vectorization           &  $77 \times 7 \times 25$ & 13475 \\
     & Fully connected layer           &  13475 & \\
     & + Layer normalization + ReLU  &  & 250 \\
    4 & Fully connected layer  &  250 & 1  \\
    \hline
                     & Output& &  1   \\

     \hline
    \hline
    \end{tabular}
\end{table}
 
\subsection{Results}

\begin{figure}[!h]
\centering
   \includegraphics[width=0.47\textwidth]{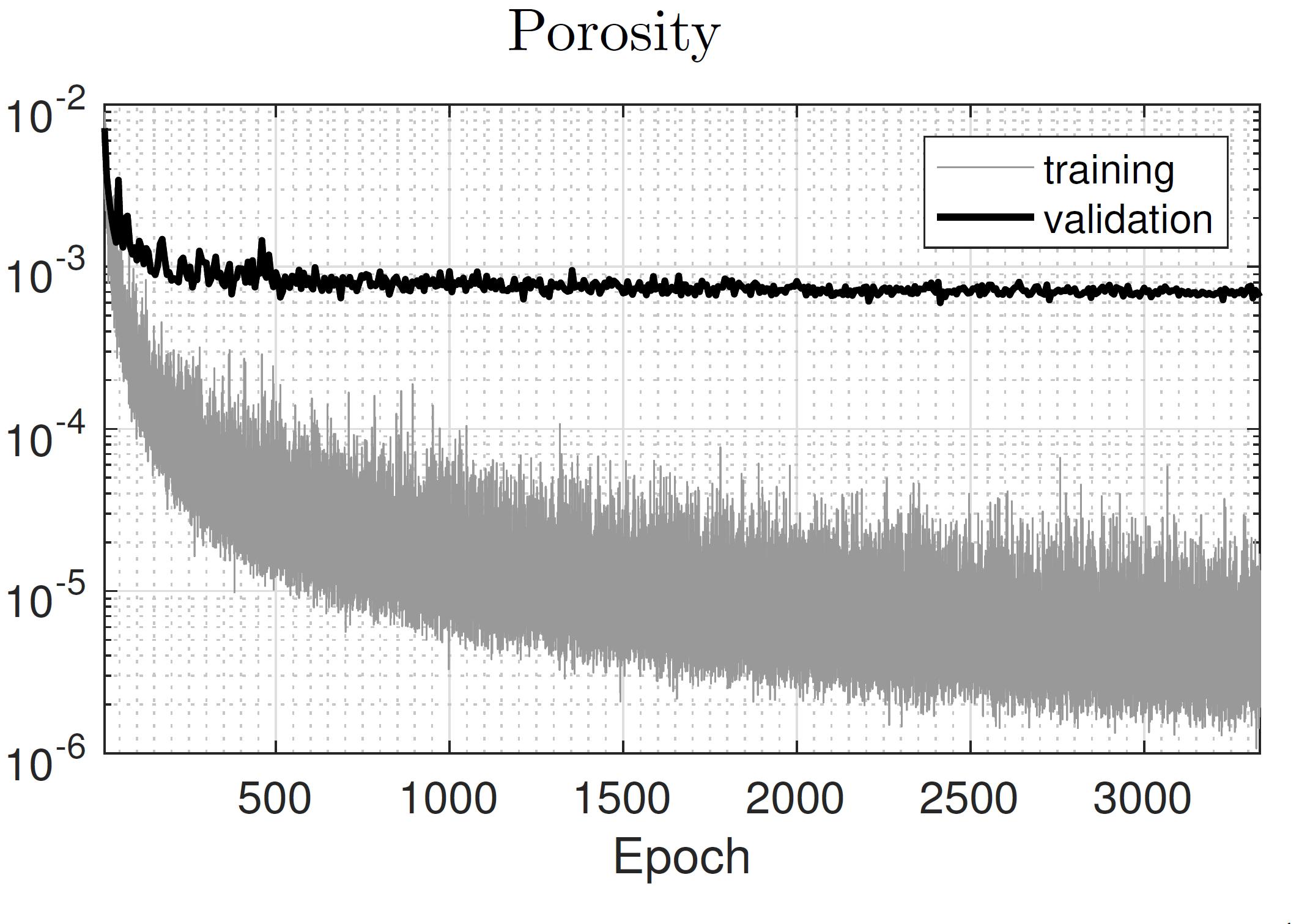}
   \includegraphics[width=0.47\textwidth]{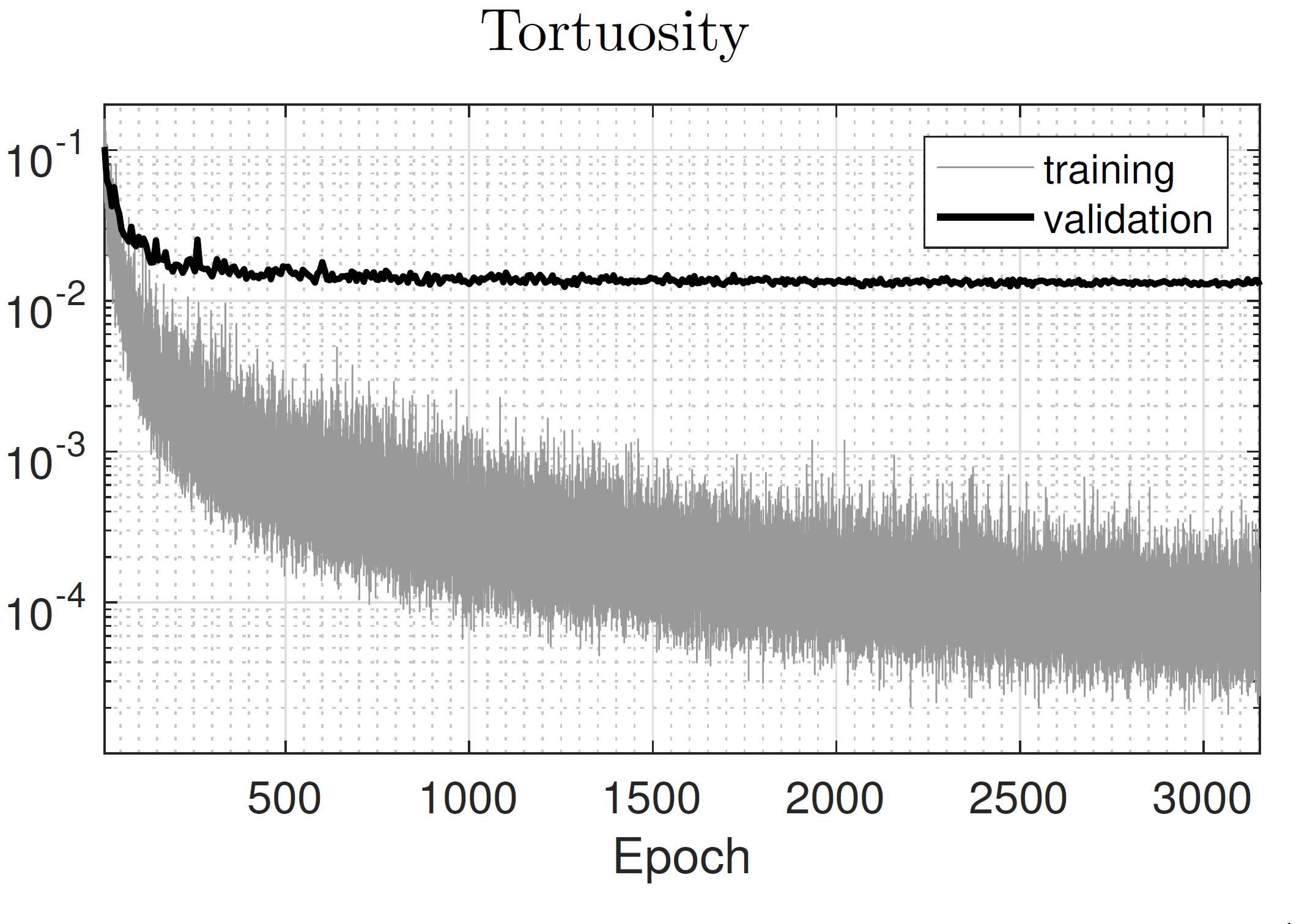}
\caption{Training and validation loss for porosity (left) and
  tortuosity (right) as function of number of epochs (full training
  cycles in stochastic optimization).}
  \label{fig:training_loss}
\end{figure}

Figure \ref{fig:training_loss} shows the loss of the training and
validation data. The loss is shown for the two unknown parameters of
interest, e.g., porosity and tortuosity. In both cases, we observe
that the network has practically reached its generalization capability
at least after 2000 full training cycles. The accuracy of the network
is affected by the marginalization over all other parameters. In
principle, the effect of the other parameters could compensate the
changes that using, for example, a different porosity would cause,
leaving the waveform of the measurement intact.

\begin{figure}[!h]
\includegraphics[width=\textwidth]{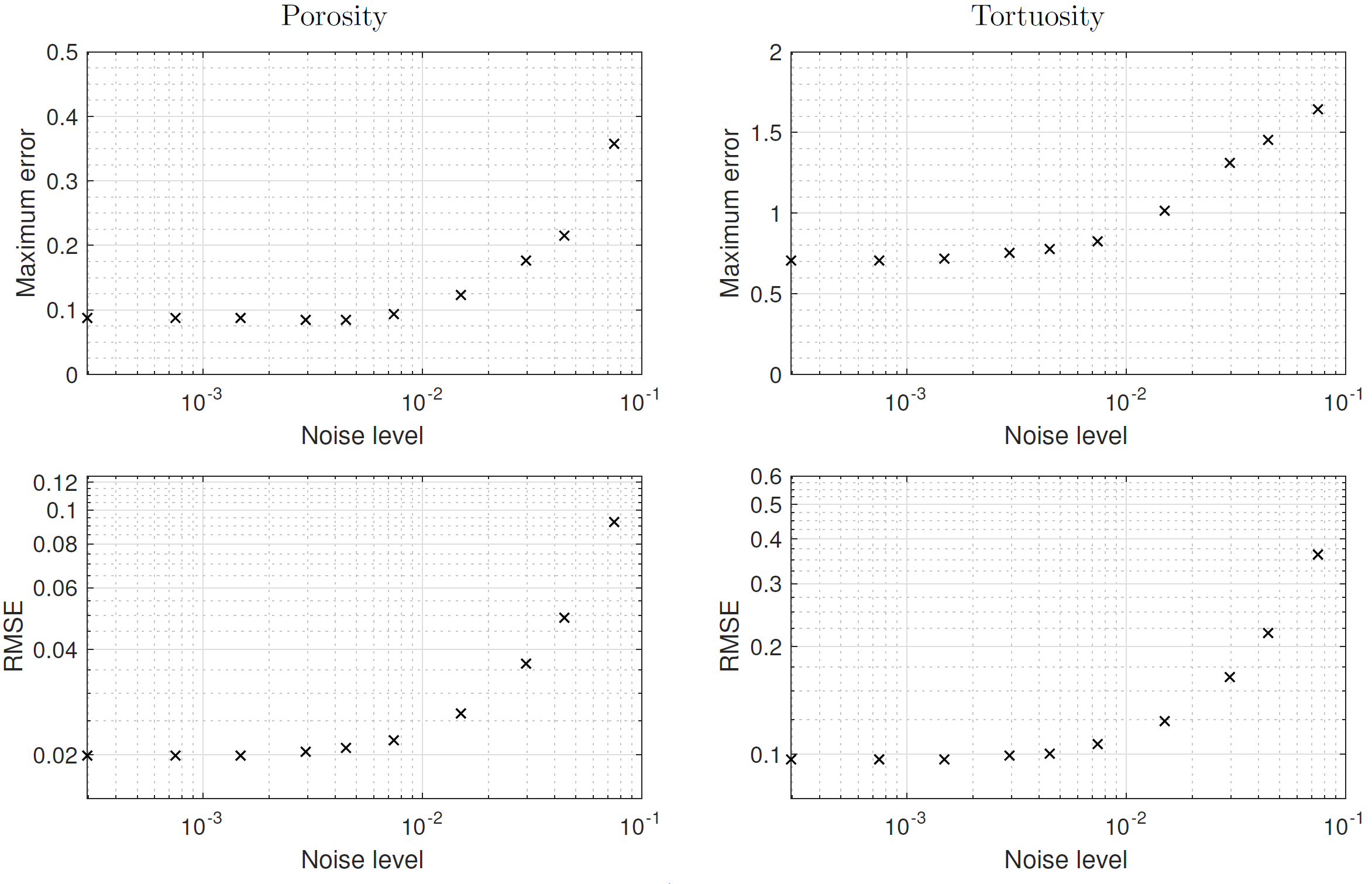}
\caption{The maximum absolute errors and the root-mean-square errors
  (RMSE) as a function of the noise level. The x-axis is the noise
  level relative to the maximum of the (noise-free) signal in the
  training set.}
  \label{fig:RMSandmaxerror}
\end{figure}

We have applied the trained network to predict porosity and tortuosity
from images of the test that are corrupted with the white noise
component with low noise level (Fig. \ref{fig:testresults_noise9}),
moderate noise level (Fig. \ref{fig:testresults_noise10}), and high
noise level (Fig. \ref{fig:testresults_noise11}).  The figures also
include error histograms.  Table \ref{tab:errorstatistics} shows
statistics of the prediction error for different noise levels.  Figure
\ref{fig:RMSandmaxerror} shows the maximum absolute error and the
root-mean-square error (the square root of Eq. (\ref{eq:loss})) as a
function of the noise level in white noise component.  The predictions
are slightly positively biased with smaller noise levels, but positive
bias is diminished with higher noise levels.  Such a behavior might be
due to the discretization error in forward models (the simulation of
training and testing data were carried out using different levels of
discretizations) which may dominate with lower observation noise
levels but becomes negligible with higher noise levels.  On the
contrary, with high levels of noise, the predictions are negatively
biased especially for larger values of porosity and tortuosity.

\begin{figure}[!h]
\includegraphics[width=\textwidth]{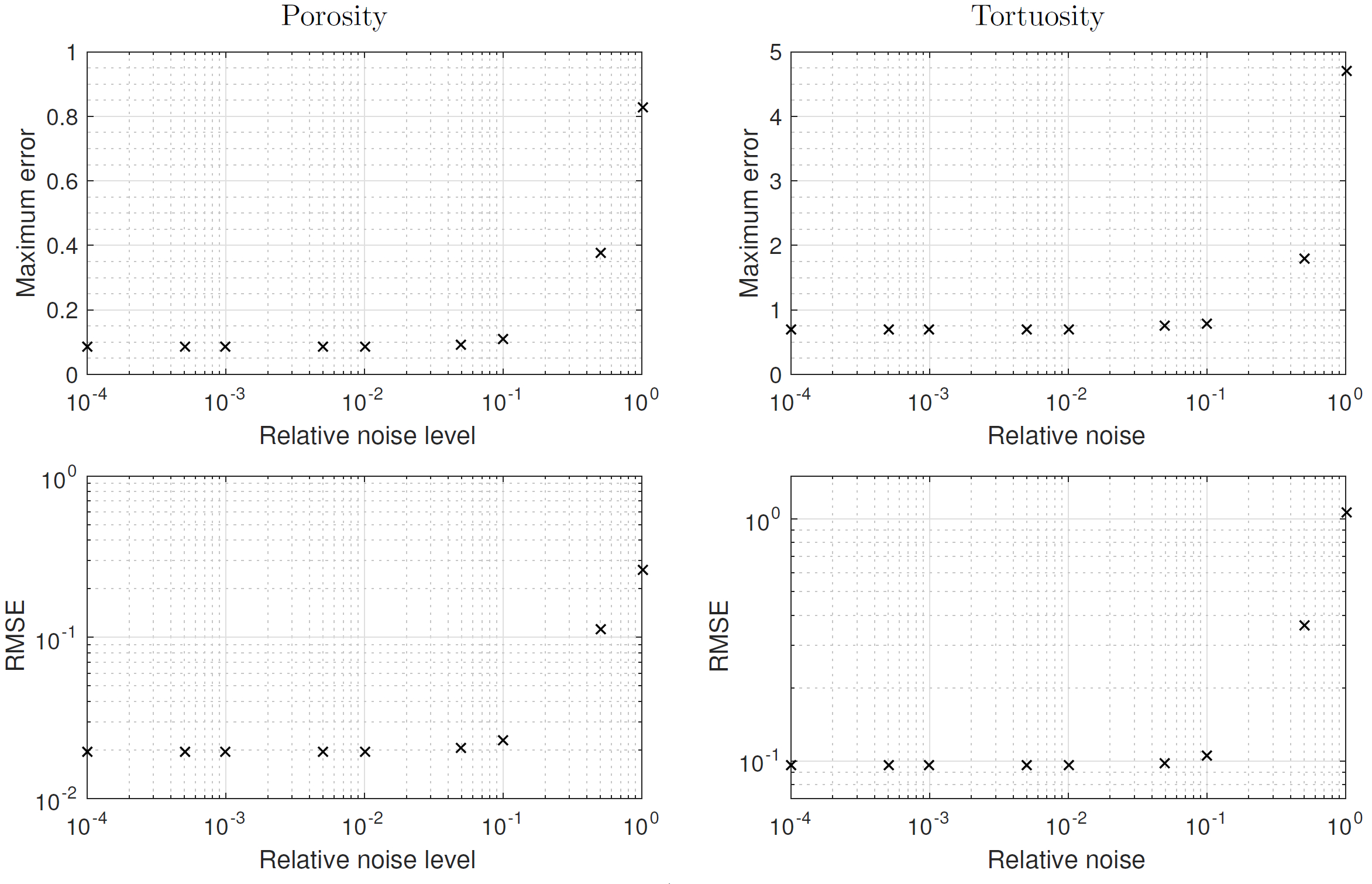}
\caption{The maximum absolute errors and the root-mean-square errors
  (RMSE) as a function of the relative noise ($B$ in Eq.
  (\ref{eq:noisemodel})).}
  \label{fig:RMSandmaxerror_rel}
\end{figure}

We have also studied the effect of the relative noise (the third term
in Eq.~(\ref{eq:noisemodel})) to the results.  Figure
\ref{fig:RMSandmaxerror_rel} shows the maximum absolute error and the
root-mean-square error as a function of the noise level in the
relative component.  As we can see, the predictions are almost
unaffected by the relative error even with significantly high noise
levels ($\sim$5-10\%).

\def\noisepicture#1#2{
\begin{figure}[!h]
\includegraphics[width=\textwidth]{Figure#2.jpg}
\caption{Predicted porosities (left) and tortuosities (middle) for the
  test data with white noise of #1 (relative to the maximum absolute
  value of the signal in the training set). Bottom row (left and
  middle) shows histograms of the prediction error (difference between
  the predicted and true values). The right column shows two examples
  of the sample images in this noise level.}
  \label{fig:testresults_noise#2}
\end{figure}
}
\noisepicture{0.8\%}{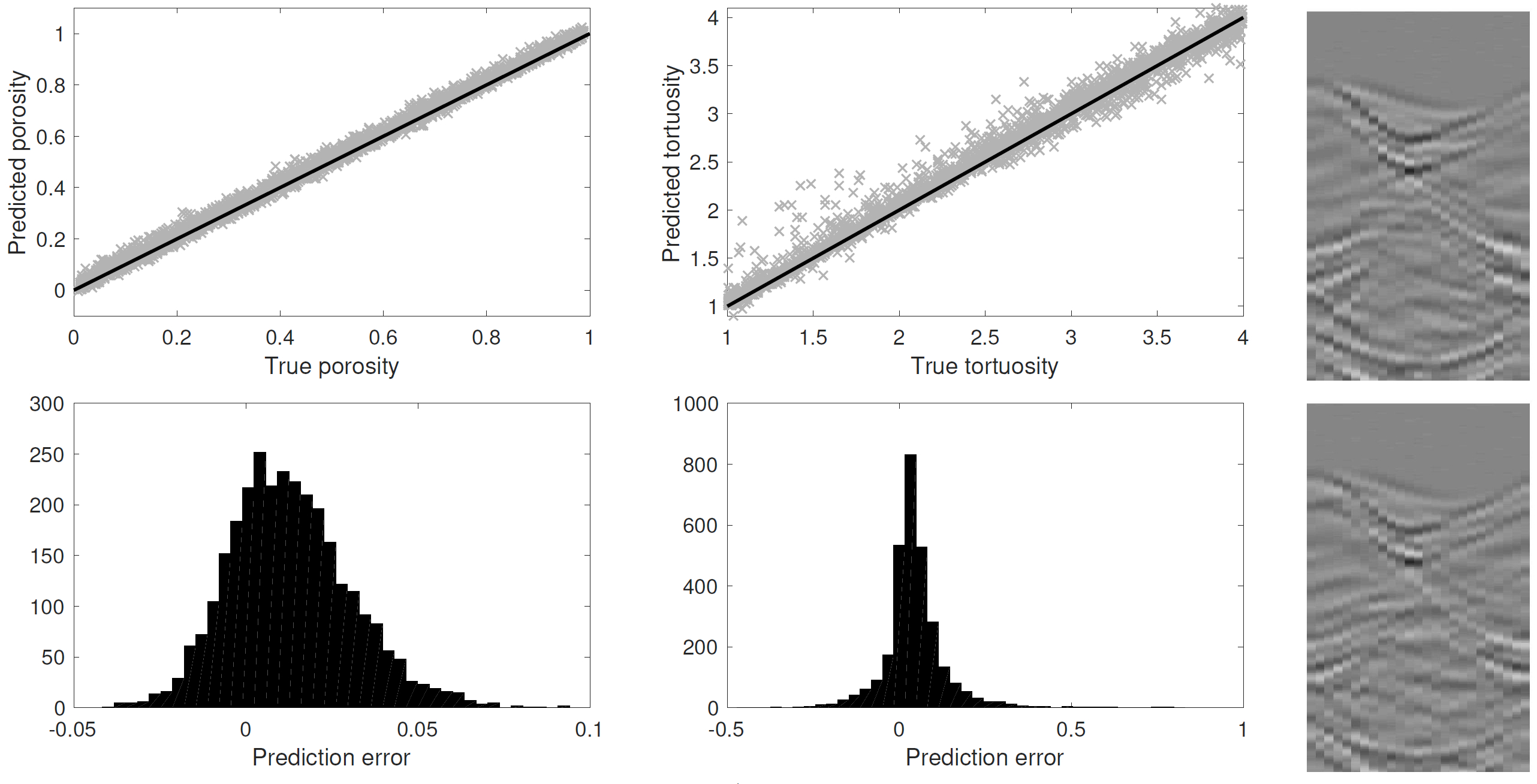}
\noisepicture{3.2\%}{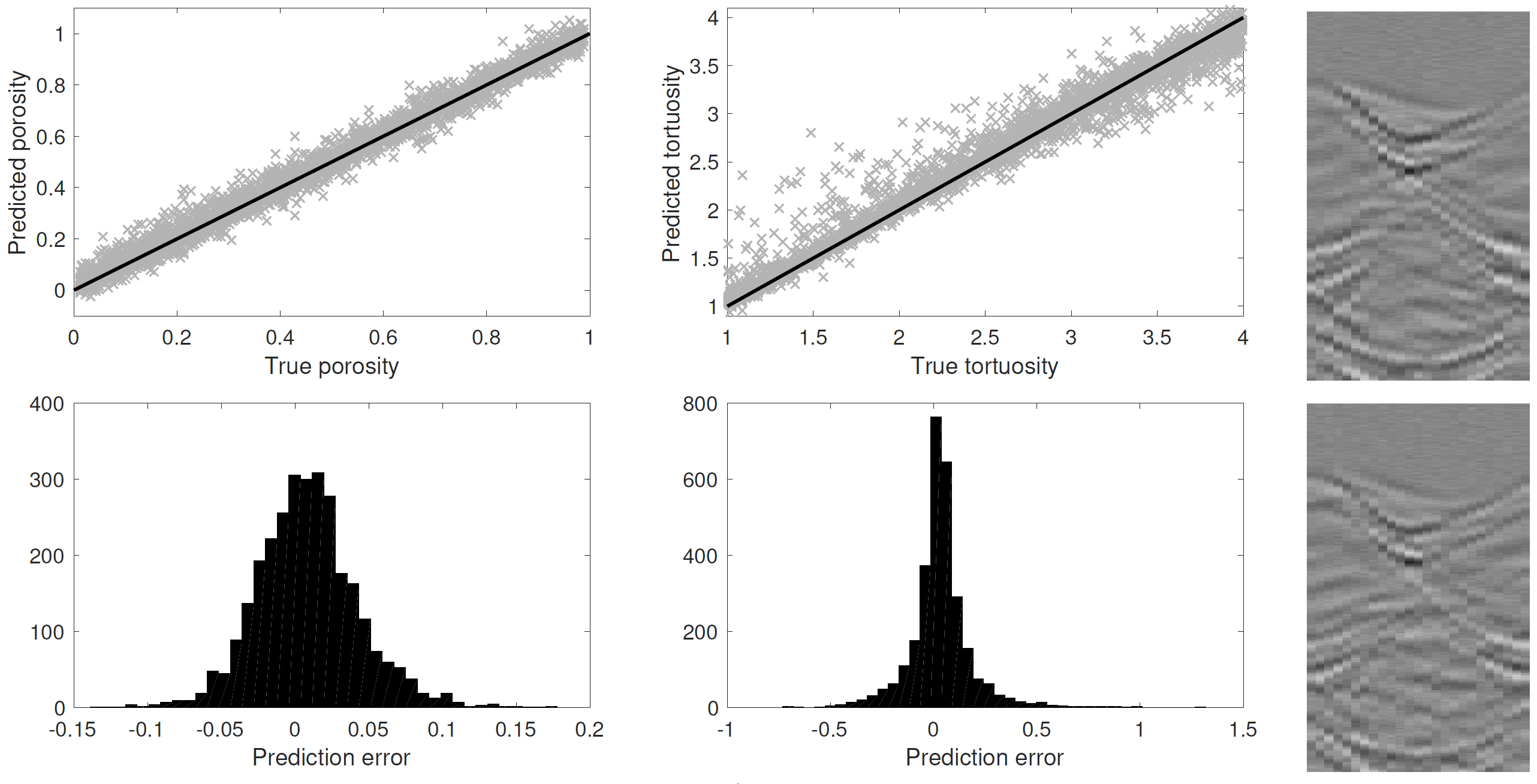}
\noisepicture{8.1\%}{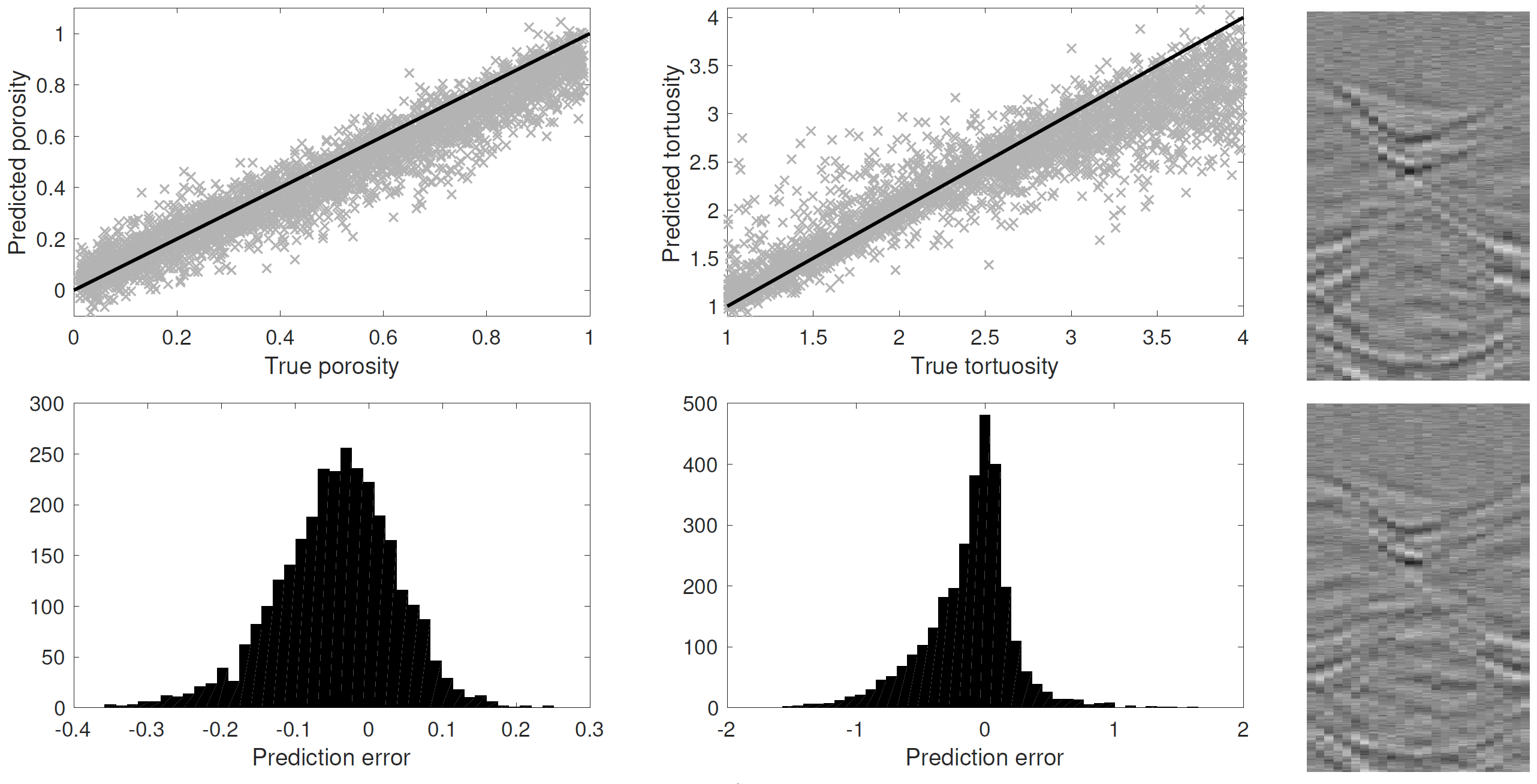}

\begin{table}[!h]
\centering
  \caption{Bias, standard deviation, kurtosis and percentiles (25\%
    and 75\%) for the errors of the predicted porosities and
    tortuosities with different noise levels.}\label{tab:errorstatistics}
  \begin{tabular}{c|ccccc|ccccc}
    \textbf{Noise}&\multicolumn{5}{c|}{\textbf{Porosity}}&\multicolumn{5}{c}{\textbf{Tortuosity}}\\
    \textbf{level} & \textbf{Bias}  & \textbf{S.D.} &
                                                            \textbf{Kurtosis}&\textbf{25\%}&\textbf{75\%}&
    \textbf{Bias}  & \textbf{S.D.} &
                                                            \textbf{Kurtosis}&\textbf{25\%}&\textbf{75\%}\\
    \hline
0 \% & 0.012 & 0.016 & 3.2 & 0.0011 & 0.023 & 0.041 & 0.087 & 14 & 0.0092 & 0.07 \\
0.16 \% & 0.012 & 0.016 & 3.2 & 0.0012 & 0.023 & 0.042 & 0.087 & 14 & 0.0097 & 0.07 \\
0.49 \% & 0.013 & 0.017 & 3.3 & 0.0013 & 0.023 & 0.044 & 0.091 & 15 & 0.0087 & 0.073 \\
0.81 \%& 0.013 & 0.018 & 3.5 & 0.00078 & 0.024 & 0.046 & 0.096 & 15 & 0.0076 & 0.075 \\
1.6 \% & 0.013 & 0.023 & 4.1 & -0.002 & 0.026 & 0.047 & 0.11 & 14 & 0.0016 & 0.08 \\
3.2 \% & 0.0087 & 0.035 & 4.4 & -0.014 & 0.028 & 0.035 & 0.16 & 11 & -0.026 & 0.085 \\
4.9 \% & -0.0029 & 0.049 & 4.2 & -0.034 & 0.027 & -0.0012 & 0.22 & 8.1 & -0.085 & 0.082 \\
8.1\% & -0.043 & 0.081 & 3.6 & -0.093 & 0.0099 & -0.12 & 0.34 & 5.1 & -0.29 & 0.062
  \end{tabular}
\end{table}

\section{CONCLUSIONS}\label{sec:conclusions}

In this paper, we proposed the use of convolutional neural networks
(CNN) for estimation of porous material parameters from synthetic
ultrasound tomography data. In the studied model, ultrasound data was
generated in a water tank into which the poroelastic material sample
was placed. A total of 26 ultrasound sensors were positioned in the
water. One of the sensors generated the source pulse while others were
used in the receiving mode. The recorded velocity data were
represented as images which were further used as an input to the CNN.
 
In the experiment, the parameter space for the porous inclusion was
assumed to be large. For example, the porosity of the inclusion was
allowed to span the interval from 1\% to 99\%. The selected parameter
space models a different type of materials.  We estimated the porosity
and tortuosity of the porous material sample while all other material
parameters were considered as nuisance parameters (see Table
\ref{tab:bounds}). Based on the results, it seems that these
parameters can be estimated with acceptable accuracy with a wide
variety of noise levels, while the nuisance parameters are
successfully marginalized. The error histograms for both porosity and
tortuosity show excellent accuracy in terms of root-mean-square error
and bias.
 
We have marginalized our inference of porosity and tortuosity over 6
other material parameters (listed in Table \ref{tab:bounds}), which
makes it possible to detect the primary porous material parameters
from the waveforms without the knowledge of the values of the nuisance
parameters, even if they can have a significant impact on the
waveforms themselves. The success in the marginalization significantly
increases the potential of the neural networks for material
characterization.
 
Future studies should include more comprehensive investigation of
model uncertainties including geometrical inaccuracies (positioning
and size of the material sample), fluid parameter changes,
viscoelastic parameters of the shell layer, material inhomogeneities,
and the sensor setup. In addition, the extension to three spatial
dimensions together with actual measurements are essential steps to
guarantee the effectiveness of the proposed method.

\subsection*{{\bf Acknowledgments}}
  This work has been supported by the strategic funding of the
  University of Eastern Finland and by the Academy of Finland (project
  250215, Finnish Centre of Excellence in Inverse Problems Research).
  This article is based upon work from COST Action DENORMS CA-15125,
  supported by COST (European Cooperation in Science and Technology).

\end{document}